\documentclass{iet-ell}

\usepackage{algorithmic}
\usepackage{algorithm}
\usepackage{svg}
\usepackage{tabularx}
\usepackage{booktabs}
\usepackage{multirow}
\usepackage{empheq}
\usepackage{array, makecell} %
\usepackage{caption}
\usepackage{subcaption}

\usepackage[nolist]{acronym}

\begin{acronym}
\acro{BS}{base station} 
\acro{UE}{user equipment} 
\acro{ZF}{zero forcing}
\acro{FoM}{figure of merit}
\acro{KPI}{key performance indicator}
\acro{RV}{random variable}
\acro{LoS}{line-of-sight}
\acro{NLoS}{non-line-of-sight}
\acro{MC}{mutual coupling}
\acro{MCM}{mutual coupling matrix}
\acro{EEP}{embedded element pattern}
\acro{AOA}{angle-of-arrival}
\acro{FD}{full-digital}
\acro{IEG}{instantaneous effective gain}
\acro{QoS}{quality-of-service}
\acro{SINR}{signal-to-interference-plus-noise ratio}
\acro{MIMO}{multiple-input-multiple-output}
\acro{SNR}{signal-to-noise ratio}
\acro{SV}{steering vector}
\acro{GSCC}{generalized spatial correlation coefficient}
\acro{ULA}{uniform linear array}
\acro{CSI}{channel state information}
\acro{NULA}{non-uniform linear array}
\acro{CDF}{cumulative distribution function}
\acro{BCC}{beamforming-channel correlation}
\newacroplural{FoM}[FoMs]{figures-of-merit}
\newacroplural{AOA}[AOAs]{angles-of-arrival}
\end{acronym}

\setcopyright{}
\ietVolume{00}
\ietYear{2023}
\ietDoi{}

\received{DD MM YYYY}
\accepted{DD MM YYYY}

\begin{document}

\title{Stochastic Phased Array Performance Indicators for Quality-of-Service-Enhanced Massive MIMO}

\author[af1]{Noud Kanters}
\orcid{0000-0001-7576-8290}
\author[af1,af2]{Andrés Alayón Glazunov}
\orcid{0000-0003-2101-4519}
\affil[af1]{Department of Electrical Engineering, University of Twente, 7500AE Enschede, Netherlands.}
\affil[af2]{Department of Science and Technology, Link\"{o}ping University, Bredgatan 33, 602 21 Norrk\"{o}ping, Sweden.}

\corresp{Email: n.b.kanters@utwente.nl}

\begin{abstract}
    In this paper, we show that the \ac{SINR} at a \ac{BS} equipped with an arbitrary physical array antenna can be expressed as a function of two fundamental \acp{FoM}: (I) the \ac{IEG}, and (II) the \ac{BCC}. These two \acp{FoM} are functions of the array antenna layout, the antenna elements, the propagation channel and the applied signal processing algorithms, and hence they are \acp{RV} in general. We illustrate that both \acp{FoM} provide essential insights for \ac{QoS}-based phased array design by investigating their statistics for \acp{BS} applying \ac{FD} \ac{ZF} beamforming. We evaluate various array designs and show that arrays with higher \acp{IEG} and a reduced probability of low \acp{BCC} can increase the ergodic sum rate and reduce the need for scheduling.
\end{abstract}

\maketitle%

\acresetall

\section{Introduction}
    Phased array antennas are a key component of \acp{BS} in multi-user wireless communication systems. Traditionally, they are configured along a uniform half-wavelength-spaced lattice to prevent grating lobes. Recently, however, communication-oriented array design \cite{oliveri2019new} has shown that unconventional layouts can enhance the \ac{UE} \ac{QoS}. Examples of considered \ac{QoS}-based \acp{KPI} are ergodic channel capacity \cite{amani2022sparse}, \ac{SINR} \cite{bencivenni2017effects, aslan2021system}, or \ac{SINR}-dependent metrics like bit error rate \cite{ge2016multi} and ergodic sum rate \cite{ge2016multi, pinchera2017antenna, amani2020multi, farsaei2020uniform}. However, understanding the physical phenomena behind array-layout-induced \ac{QoS} improvements is complicated, as is illustrated by, for instance, conflicting statements on whether \ac{MC} enhances or deteriorates ergodic channel capacity; see, e.g., \cite{chen2018review} and references therein. Typically, different assumptions are made regarding the number and type of antenna elements, the number of served \acp{UE}, and the propagation channel model. Moreover, conventional channel normalization may partially hide the impact of the element type, its element pattern, and impedance matching. This hinders the straightforward comparison of the proposed array designs. Conventional phased array \acp{KPI} like sidelobe-level and beamwidth provide limited insight into how an array will perform in a multi-user system, especially in channels with a \ac{NLoS} component. Hence, generalized \acp{KPI} incorporating the effects of the array, the channel and signal processing are needed. In this work, we derive such \acp{KPI}. Specifically, we show that the \ac{SINR} in single-cell systems solely depends on the transmit powers and two \acp{RV}: (I) the \ac{IEG} and (II) the \ac{BCC}. This result is obtained by normalizing the \ac{BS}-\ac{UE} channels based on a \ac{BS} reference array. Subsequently, we illustrate how the array design can affect the statistics of these two \acp{RV}, and with that also of \ac{QoS}-based array design objectives. To this end, we consider \acp{BS} equipped with various linear arrays applying \ac{FD} \ac{ZF} beamforming, both with and without user scheduling, and we analyse how the statistics of the two \acp{RV} affect the achieved \ac{SINR} and ergodic sum rate.
    
\section{Massive MIMO System Model} 
    
        Let's consider a single-cell massive \ac{MIMO} system comprising a \ac{BS} serving $K$ \acp{UE}. Each \ac{UE} has a single antenna element, whereas the \ac{BS} has an $N$-element phased array antenna. The narrowband uplink received signal $\mathbf{y}^\mathrm{UL}\in\mathbb{C}^N$ is defined as in, e.g., \cite{bjornson2017massive} and reads 
        \begin{equation}
        \label{eq:y_UL}
            \mathbf{y}^\mathrm{UL} = \sum_{k=1}^{K} \sqrt{p_{k}}  \mathbf{h}_{k} x_{k} + \sigma_\mathrm{UL} \mathbf{n},
        \end{equation}
        where $\mathbf{h}_k\in\mathbb{C}^N$, $p_k$ and $x_k\sim\mathcal{N}_{\mathbb{C}}(0,1)$ represent the \ac{BS}-\ac{UE} channel vector, the transmit power and the data signal for the $k$\textsuperscript{th} \ac{UE}, respectively. Moreover, $\mathbf{n}\sim\mathcal{N}_\mathbb{C}(\mathbf{0}_N,\mathbf{I}_N)$ is the receiver noise vector and $\sigma_\mathrm{UL}^2$ the noise power. Assuming the \ac{BS} applies linear receive combining using combining matrix $\mathbf{W}\in\mathbb{C}^{N\times K}=\begin{bmatrix} \mathbf{w}_{1} &\cdots & \mathbf{w}_{k}\end{bmatrix}$, it follows that the instantaneous uplink \ac{SINR} for \ac{UE} $k$ equals
        \begin{equation}
        \label{eq:SINR_UL_gnrc}
            \mathrm{SINR}^{\mathrm{UL}}_{k} = \frac{p_{k}|\mathbf{w}_{k}^H \mathbf{h}_{k}|^2}{\underbrace{\vphantom{\sum_{\substack{l=1, l\neq j}}^{L}} \sum_{\substack{i=1, i\neq k}}^{K} p_{i} |\mathbf{w}_{k}^H \mathbf{h}_{i}|^2}_{\text{Intra-Cell Interference}} + \underbrace{\vphantom{\sum_{\substack{l=1, l\neq j}}^{L}}\sigma_\mathrm{UL}^2 \lVert \mathbf{w}_{k} \rVert^2}_{\text{Noise}}}.
        \end{equation}

\section{SINR-Dependent QoS-based Array Design Single-Cell Systems}
\label{sec:coeff}

    In this section, we present the channel normalization technique assumed in the remainder of this paper. We use this to derive a novel expression for the \ac{SINR} in single-cell systems. The result applies to arbitrary array layouts, arbitrary linear combining algorithms, and arbitrary channel models.
  

     Before computing the \ac{SINR} in \eqref{eq:SINR_UL_gnrc}, some form of norm-based normalization is generally applied to $\mathbf{h}_1,\dots,\mathbf{h}_K$ to have control over the \ac{SNR}. Conventional normalization approaches are $\lVert\mathbf{h}_k\rVert=\sqrt{N} \; \forall \; k=1,\dots,K$ or $\lVert \mathbf{H} \rVert_F=\sqrt{NK}$, where $\mathbf{H}=\begin{bmatrix} \mathbf{h}_1 \cdots \mathbf{h}_K \end{bmatrix}$.
     However, when designing physical array antennas based on a \ac{QoS}-metric like the \ac{SINR}, the impact of the antennas is typically embedded in the channel vectors $\mathbf{h}_1,\dots,\mathrm{h}_K$, see, e.g., \cite{amani2022sparse, ge2016multi}. Applying these conventional normalization techniques cancels out essential information regarding the relation between, on the one hand, deterministic array aspects like \ac{MC} and impedance matching and, on the other hand, the stochastic propagation environment in which the array is deployed. Consequently, assessing the performance of various array designs deployed within a certain channel or of a specific array deployed in different channels is not straightforward. To circumvent this problem, we propose normalizing the \ac{BS}-\ac{UE} channels relative to a reference array rather than in an absolute sense. Hence, the normalized channel between the \ac{BS} and \ac{UE} $k$ is defined as
    \begin{equation}
    \label{eq:normalization}
        \mathbf{h}_k = \sqrt{N_\mathrm{ref}} \frac{  \tilde{\mathbf{h}}_k }{ \lVert\tilde{\mathbf{h}}_k^\mathrm{ref}\rVert},
    \end{equation}
    where we use $\tilde{\mathbf{h}}$ and $\mathbf{h}$ to differentiate between non-normalized channels and their normalized counterparts as used in \eqref{eq:SINR_UL_gnrc}, respectively. $\tilde{\mathbf{h}}_k^\mathrm{ref}$ represents the \ac{BS}-\ac{UE} channel that would be observed if the \ac{BS} array of interest were replaced by the reference array while leaving the propagation channel (defined by parameters like, e.g, \acp{AOA}, complex path gains and Rice factor) unchanged. $N_\mathrm{ref}$ is the number of elements in the reference array. The reference array does not need the same number of elements as the array of interest. Note that the normalized channel between a \ac{UE} and the reference array by definition satisfies $\lVert \mathbf{h}_k^\mathrm{ref}\rVert=\sqrt{N_\mathrm{ref}}$. Although not required, we consider reference arrays composed of isotropic elements in this work. 
        

    Assuming that $p_1=\dots=p_K=P_\mathrm{UL}/N_\mathrm{ref}$ and applying \eqref{eq:normalization}, it follows that \eqref{eq:SINR_UL_gnrc} can be written as
    \begin{equation}
    \label{eq:SINR_UL_KPIs}
        \mathrm{SINR}^{\mathrm{UL}}_{k} =  \frac{ P_\mathrm{UL} \, G^\mathrm{ie}_k \, |\omega_{kk}|^2}{\underbrace{P_\mathrm{UL} \sum\limits_{\substack{i=1, i\neq k}}^{K} G^\mathrm{ie}_i \, |\omega_{ki}|^2 }_{\text{Intra-Cell Interference}} +  \underbrace{\vphantom{\sum\limits_{\substack{i=1, i\neq k}}^{K}}1}_{\text{Noise}}},
    \end{equation}
    where we have assumed without loss of generality that $\sigma_\mathrm{UL}=1$, and where we've introduced the complex-valued \ac{BCC} coefficient $\omega_{ki}$ and the \ac{IEG} $G^\mathrm{ie}_i$. Here, $\omega_{ki}$ is defined as 
    \begin{equation}
    \label{eq:omega}
        \omega_{ki} = \frac{\mathbf{w}_{k}^H \mathbf{h}_{i}}{\lVert\mathbf{w}_{k}\rVert \lVert\mathbf{h}_{i}\rVert},
    \end{equation}
    which satisfies $0 \leq |\omega_{ki}|^2 \leq 1$ for $i\in\{1,\dots,K\}$. 
    From \eqref{eq:omega}, it follows that $|\mathbf{w}_{k}^H \mathbf{h}_{i}|^2=|\omega_{ki}|^2 \lVert \mathbf{w}_{k}^H \rVert^2 \lVert \mathbf{h}_{i} \rVert^2$. This is substituted in \eqref{eq:SINR_UL_gnrc}, whereupon we have used that $p_i \lVert \mathbf{h}_{i} \rVert^2 = \frac{P_\mathrm{UL}}{N_\mathrm{ref}} \lVert \sqrt{N_\mathrm{ref}} \frac{\tilde{\mathbf{h}}_{i}}{\lVert\tilde{\mathbf{h}}_{i}^\mathrm{ref}\rVert} \rVert^2 = P_\mathrm{UL} \frac{\lVert \tilde{\mathbf{h}}_{i} \rVert^2}{\lVert \tilde{\mathbf{h}}^\mathrm{ref}_{i} \rVert^2} = P_\mathrm{UL} \frac{\lVert \mathbf{h}_{i} \rVert^2}{\lVert \mathbf{h}^\mathrm{ref}_{i} \rVert^2} =P_\mathrm{UL}G^\mathrm{ie}_i$, where the last steps follow from \eqref{eq:normalization} and from the definition of the \ac{IEG}, i.e.,
    \begin{equation}
    \label{eq:EAG}
       G^\mathrm{ie}_i = \frac{\lVert \mathbf{h}_i \rVert^2}{\lVert \mathbf{h}_i^\mathrm{ref} \rVert^2}.
    \end{equation} 
    Hence, in the definition of the \ac{IEG}, the numerator represents the instantaneous channel gain observed at the physical \ac{BS} array under consideration, whereas the denominator represents, for the same \ac{UE} and the same propagation channel, the instantaneous channel gain observed at the reference array. Therefore, the \ac{IEG} measures the channel gain of an array of physical elements relative to an isotropic array with no \ac{MC}. It is worthwhile to note that an expression similar to \eqref{eq:SINR_UL_KPIs} is obtained for the downlink when assuming that the downlink transmit power is defined as $p_{i}\lVert\mathbf{w}_i\rVert^2=P_\mathrm{DL}/N_\mathrm{ref}$ for all $i\in\{1,\dots, K\}$. In this case, expressions for uplink and downlink \ac{SINR} are equivalent if $P_\mathrm{UL}=P_\mathrm{DL}$. For the sake of conciseness, we only focus on the uplink.
     
    The principle behind \ac{SINR}-dependent \ac{QoS}-based array layout design in single-cell systems can be understood from \eqref{eq:SINR_UL_KPIs}. The stochastic propagation channel, the deterministic \ac{BS} array antenna, and the applied signal processing algorithms (e.g., user scheduling and beamforming) jointly determine the statistics of the coefficients $G^\mathrm{ie}_i$ and $|\omega_{ki}|^2$, $i\in\{1,\dots, K\}$. Both coefficients are \acp{RV} in general, and consequently, $\mathrm{SINR}_{k}^\mathrm{UL}$ is an \ac{RV} as well. Through proper design of the array antenna, the probability distributions of $G^\mathrm{ie}_i$ and $|\omega_{ki}|^2$ can be shaped to optimize the design objective, which is typically a specific statistic of (a function of) $\mathrm{SINR}_{k}^\mathrm{UL}$.        

\section{Channel Model and Signal Processing}

    The theory presented in this paper applies to arbitrary channel models. However, we limit ourselves to pure \ac{LoS} far-field channels for conciseness. Furthermore, we assume that all antenna elements are purely vertically polarized. The \acp{UE} use isotropic antennas, whereas the \ac{BS} uses a physical array antenna. Hence, the channel between \ac{UE} $k$ and the \ac{BS} can be represented as \cite{friedlander2017antenna} 
    \begin{subequations}
    \label{eq:a_analytic}
    \begin{align}
        \tilde{\mathbf{h}}_k & =\mathbf{a}(\phi_k,\theta_k) \\
        & = \mathbf{g}(\phi_k,\theta_k)\odot\mathbf{a}_\mathrm{isotropic}(\phi_k,\theta_k) \label{eq:subeq1}\\
        & = \mathbf{C}_\mathrm{oc}(\phi_k,\theta_k) \; \mathbf{a}_\mathrm{isotropic}(\phi_k,\theta_k) \label{eq:subeq2},
    \end{align}
    \end{subequations}
     where $\mathbf{a}(\phi_k,\theta_k)$ is the analytic array manifold \cite{friedlander2017antenna} at $(\phi_k,\theta_k)$, the azimuth and elevation \acp{AOA} for 
     \ac{UE} $k$. The tilde in $\tilde{\mathbf{h}}_k$ indicates that this is the channel vector before normalization according to \eqref{eq:normalization}. In \eqref{eq:subeq1} and \eqref{eq:subeq2}, $\mathbf{a}_\mathrm{isotropic}(\phi,\theta)\in\mathbb{C}^N$ represents the array's \ac{SV} defined as 
    \begin{equation}
    \label{eq:a_isotropic}
     \mathbf{a}_\mathrm{isotropic}(\phi,\theta) =
        \begin{bmatrix}
            \mathrm{exp}(-j\frac{2\pi}{\lambda}\mathbf{r}_1 \cdot \mathbf{u}(\phi,\theta)) \\ \vdots \\ \mathrm{exp}(-j\frac{2\pi}{\lambda}\mathbf{r}_N \cdot \mathbf{u}(\phi,\theta))
        \end{bmatrix},
    \end{equation} 
    where $\lambda$ is the wavelength, $\mathbf{r}_n\in\mathbb{R}^3$ represents the position of element $n$ in space in Cartesian coordinates, and $\mathbf{u}(\phi,\theta)=\begin{bmatrix}
        \cos(\phi)\cos(\theta), \, \sin(\phi)\cos(\theta), \, \sin(\theta)
    \end{bmatrix}^T$ is a unit vector in the direction of ($\phi$, $\theta$). The impact of the physical antenna elements is modelled using \acp{EEP} \cite{pozar1994active} in \eqref{eq:subeq1} and using a \ac{MCM} in \eqref{eq:subeq2}. Specifically, the vector $\mathbf{g}(\phi,\theta)\in\mathbb{C}^N$ is defined as $\mathbf{g}(\phi,\theta)=\begin{bmatrix}
        g_{1}(\phi,\theta) \cdots g_{N}(\phi,\theta)
    \end{bmatrix}^T$, where $g_{n}(\phi,\theta)$ represents the \ac{EEP} of element $n$, whereas $\mathbf{C}_\mathrm{oc}(\phi,\theta)$ is the direction-dependent \ac{MCM} defined as \cite{friedlander2017antenna} 
    \begin{equation}
    \label{eq:MCM}
        \mathbf{C}_\mathrm{oc}(\phi,\theta) = \mathbf{Z}_\mathrm{L}(\mathbf{Z}+\mathbf{Z}_\mathrm{L})^{-1} \mathbf{G}_\mathrm{oc}(\phi,\theta).
    \end{equation}
    Here, $\mathbf{Z}\in\mathbb{C}^{N\times N}$ is the mutual impedance matrix,  $\mathbf{Z}_\mathrm{L}\in\mathbb{C}^{N\times N}$ is the load impedance matrix, and $\mathbf{G}_\mathrm{oc}(\phi,\theta) \in \mathbb{C}^{N\times N}$ is a diagonal matrix defined as
    \begin{equation}
    \label{eq:G_oc}
        \mathbf{G}_\mathrm{oc}(\phi,\theta) = \mathrm{diag}\big(\big[g_{\mathrm{oc},1}(\phi,\theta), \dots, g_{\mathrm{oc},N}(\phi,\theta) \big] \big),
    \end{equation}
    where $g_{\mathrm{oc},n}(\phi,\theta)$ is the open-circuit element pattern of element $n$, i.e., the pattern of element $n$ when embedded in the array with all other elements open-circuited. 

    A convenient simplification of \eqref{eq:MCM} exists for \ac{BS} arrays composed of thin dipoles with inter-element spacings larger than quarter-wavelength \cite{craeye2011review}. For these arrays, dipole elements behave as minimum scattering antennas, meaning their open-circuit patterns are approximately equivalent to the isolated element patterns, see, e.g., \cite{clerckx2007impact}. Since isolated dipoles are omni-directional in the plane orthogonal to the dipole axis, it follows that for an array of identical dipole elements oriented vertically in a horizontal plane, $\mathbf{G}_\mathrm{iso}(\phi,\theta)\propto\mathbf{I}_N$. Hence, the \ac{MCM} \eqref{eq:MCM} becomes $\mathbf{C}_\mathrm{iso} \propto \mathbf{Z}_\mathrm{L}(\mathbf{Z}+\mathbf{Z}_\mathrm{L})^{-1}$ and is therefore direction-independent. For a \ac{BS} array of isotropic elements, we can set $\mathbf{g}(\phi,\theta)=\mathbf{1}_N$ in \eqref{eq:subeq1} and define
     \begin{equation}
    \label{eq:h_isotropic}
         \tilde{\mathbf{h}}_\mathrm{isotropic} = \tilde{\mathbf{h}}(\mathbf{g}(\phi,\theta)=\mathbf{1}_N) = \mathbf{a}_\mathrm{isotropic}(\phi,\theta),
    \end{equation} 
    where we have omitted the subscript $k$ for ease of notation. 
        


\begin{figure}
            \centering
            \def\svgwidth{0.99\columnwidth}
            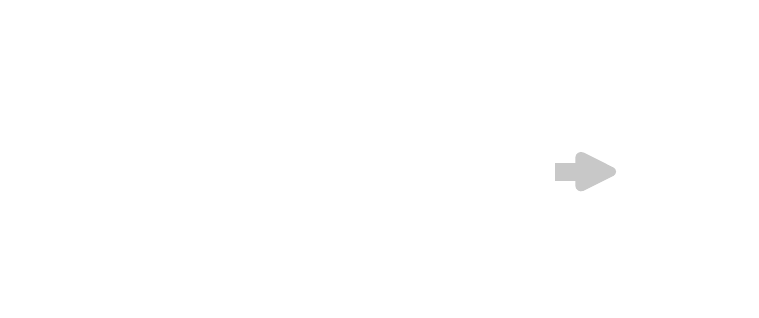
            \caption{Ergodic sum rate-based array design for \ac{FD} \ac{ZF} single-cell systems.}
            \label{fig:alpha2R}
        \end{figure}

    Assuming perfect \ac{CSI} is available at the \ac{BS}, the \ac{ZF} combining matrix is computed as $\mathbf{W} = \mathbf{H}(\mathbf{H}^{H} \mathbf{H})^{-1}$ \cite{marzetta2016fundamentals, bjornson2017massive}. By definition, the \ac{ZF} combining vector for \ac{UE} $k$, $\mathbf{w}_{k}$, is orthogonal to the (intra-cell) interference subspace $I_k$, i.e., the vector space spanned by the $K-1$ channel vectors $\mathbf{h}_i$, $i\in\{1,\dots,K\}\setminus k$ \cite{bjornson2014optimal}. Hence, for \ac{ZF} combining, the beamforming-channel correlation satisfies $|\omega_{ki}|^2=0$ for all $i\in\{1,\dots,K\}\setminus k$, meaning the interference term in the denominator of \eqref{eq:SINR_UL_KPIs} vanishes. Moreover, it follows from \cite{lin1982spatial} that
    \begin{equation}
    \label{eq:omega_ZF}
        |\omega_{kk}|^2 \Big|_\mathrm{ZF} = 1- |\cos(\bar{\gamma}_{kk})|^2 = 1-|\bar{\alpha}_{kk}|^2,
    \end{equation}
    where $\bar{\gamma}_{kk}$ represents the generalized angle between $\mathbf{h}_k$ and its projection on $I_k$, and where $|\bar{\alpha}_{kk}|^2$ represents the corresponding \ac{GSCC} which can be computed using, e.g., the Gram-Schmidt procedure \cite{lin1982spatial}. Hence, for \ac{ZF} combining, \eqref{eq:SINR_UL_KPIs} reduces to
    \begin{equation}
    \label{eq:SINR_ZF_SC}
         \mathrm{SINR}_{k}^{\mathrm{UL}}|_{\mathrm{ZF}} =  P_\mathrm{UL} \,  G^\mathrm{ie}_k \, (1-|\bar{\alpha}_{kk}|^2).
    \end{equation}
    Fig.~\ref{fig:alpha2R} visualizes the array design procedure for ergodic sum rate-based design in the case of \acp{BS} applying \ac{FD} \ac{ZF} combining. 
        
    \section{Numerical Simulations and Simulation Parameters}

        \setlength{\tabcolsep}{4pt} 
        \begin{table}
            \centering	
            \caption{Antenna Elements}
            \label{tab:elements}
            \centering\renewcommand\cellalign{lc}
            \def\arraystretch{0.8}
            \begin{tabularx}{\columnwidth}{@{} r l c X @{}}
                \toprule
                    & \textbf{Element} & \textbf{Manifold} & \textbf{Element Pattern} \\
                \midrule
                   \textbf{1} & Isotropic & \eqref{eq:a_isotropic} & - \\
                   \textbf{2} & $\frac{\lambda}{2}$-Dipole & \eqref{eq:subeq2} & $g_{\mathrm{oc},n}(\phi,\theta) \propto \cfrac{\sin(\frac{\pi}{2}\sin(\theta))}{\cos(\theta)}$ \cite{balanis2015antenna} \\             
                   \textbf{3} & Cosine & \eqref{eq:subeq1} & $g_n(\phi,\theta) \propto \begin{cases}
                       \cos(\phi)\cos(\theta) & |\phi| \leq 90^\circ \\
                       0 & \text{otherwise.}
                   \end{cases}$ \\
                \bottomrule
            \end{tabularx}{}
            \vspace{-5pt}
        \end{table} 
        
        We focus on \acp{BS} applying \ac{FD} \ac{ZF} combining. Since \ac{ZF} causes low \acp{SINR} in the case of highly correlated \ac{BS}-\ac{UE} channels \cite{farsaei2020uniform}, we consider scenarios without and with user scheduling. Specifically, we apply user dropping according to \cite{yang2017massive}. A correlation threshold of $\frac{|\mathbf{h}_i^H\mathbf{h}_j|}{\lVert\mathbf{h}_i\rVert\lVert\mathbf{h}_j\rVert}\leq 0.45$ is considered in the scenario with dropping. The considered array antenna elements are presented in Table~\ref{tab:elements}. The reference array for computing the \ac{IEG} is a $\lambda/2$-spaced \ac{ULA} composed of isotropic elements. Furthermore, we consider two arrays composed of vertically oriented half-wave dipoles: a \ac{ULA} and a \ac{NULA}. The Tchebyshev parametrization of \cite{pinchera2017antenna} determines the spatial configuration of the latter.
        For both dipole arrays, each dipole is terminated in the complex conjugate of its self impedance $Z_s$ such that $\mathbf{Z}_\mathrm{L}=Z_s^*\mathbf{I}_N$ in \eqref{eq:MCM}. The mutual and self impedances are defined in, e.g., \cite{balanis2015antenna}. Finally, we consider the same \ac{NULA} but with synthetic cosine elements; see Table~\ref{tab:elements}. These elements have a directive element pattern and could thus represent, for instance, patch antennas. For all arrays, except for the reference array, we consider two (average) inter-element spacings: $d_\mathrm{avg}=\lambda/2$ and $d_\mathrm{avg}=2\lambda$. All arrays are composed of $N = N_\mathrm{ref}=32$ elements. We scale the element patterns by a factor $\gamma$ such that the integrals of their received gain patterns in the absence of \ac{MC} are equal. Hence, for the dipoles, we apply $\int_{-\pi}^{\pi}\int_{-\pi/2}^{\pi/2} | \gamma \frac{Z_s^*}{Z_s+Z_s^*} g_{\mathrm{oc},n}(\phi,\theta)|^2 \cos(\theta) d\theta d\phi = 4\pi$, ultimately resulting in the well-known 2.15 dBi gain in the horizontal plane \cite{balanis2015antenna}. Note that the impedance ratio appearing in the integral can alternatively be taken into account by introducing a factor $\frac{Z_s + Z_s^*}{Z_s^*} $ in the definition of the \ac{MCM}, see, e.g., \cite{clerckx2007impact}. For the cosine elements, we apply $\int_{-\pi}^{\pi}\int_{-\pi/2}^{\pi/2} | \gamma g_{n}(\phi,\theta)|^2 \cos(\theta) d\theta d\phi = 4\pi$. We consider a 2-dimensional horizontal geometry
        with a \ac{BS} serving $K=8$ \acp{UE}, which are uniformly distributed over a 120$^\circ$ sector. The azimuth and elevation AOAs are defined as $\phi_k\sim U(-60^\circ,60^\circ)$ and $\theta_k=0^\circ$, $k=1,\dots,K$, respectively.
        We simulate $10^4$ realizations with \ac{BS}-\ac{UE} channels modelled as \eqref{eq:a_analytic}, or as \eqref{eq:h_isotropic} for the isotropic reference array.

\section{Results}    

     \begin{figure*}
         \centering
         \begin{subfigure}[b]{\textwidth}
             \centering
             \includegraphics[width=\textwidth]{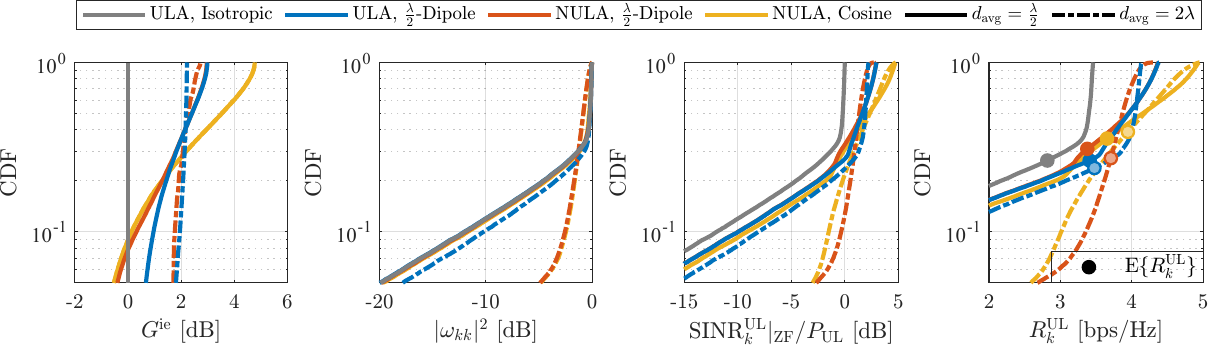}
             \caption{Without user dropping}
             \label{fig:rslt_no_dropping}
             \vspace{-3pt}
         \end{subfigure}
         \vspace{3pt}
         \\
         \begin{subfigure}[b]{\textwidth}
             \centering
             \includegraphics[width=\textwidth]{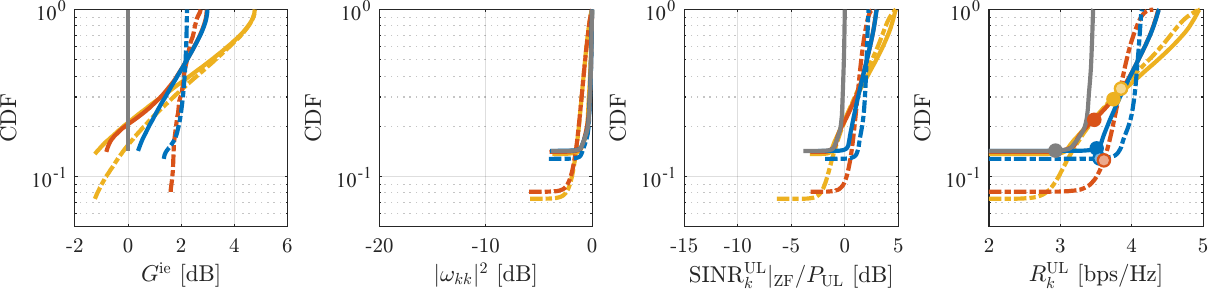}
             \caption{With user dropping}
             \label{fig:rslt_dropping}
             \vspace{-3pt}
         \end{subfigure}
            \caption{From left to right: CDFs of (I) IEG $G^\mathrm{ei}_k$ \eqref{eq:EAG}; (II) \ac{ZF} \ac{BCC} $|\omega_{kk}|^2\big|_\mathrm{ZF}$ \eqref{eq:omega_ZF}; (III) $\mathrm{SINR}^\mathrm{UL}_k|_\mathrm{ZF}/P_\mathrm{UL}$ \eqref{eq:SINR_ZF_SC}; and (IV) \ac{UE} rate $R_k^\mathrm{UL}=\log_2(1+\mathrm{SINR}_k^\mathrm{UL}|_\mathrm{ZF})$ at $P_\mathrm{UL}=10$ dB.}
            \label{fig:results}
            \vspace{-5pt}
    \end{figure*}
           
    Results are presented in Fig.~\ref{fig:rslt_no_dropping} and Fig.~\ref{fig:rslt_dropping} for the scenarios without and with user dropping, respectively. They are discussed below. 


    The first subplots of Fig.~\ref{fig:rslt_no_dropping} and Fig.~\ref{fig:rslt_dropping}  show, in dB-scale, the \acp{CDF} of the \ac{IEG} $G^\mathrm{ie}_k$ \eqref{eq:EAG}. The percentile at which a graph ends in Fig.~\ref{fig:rslt_dropping} indicates the probability of a user being dropped. Clearly, this probability is the lowest for the \acp{NULA} with $d_\mathrm{avg}=2\lambda$. Looking at the served (i.e., non-dropped) \acp{UE} alone, it can be seen that user dropping has a negligible impact on the statistics of the \ac{IEG} $G^\mathrm{ie}_k$. Furthermore, it is observed that the isotropic reference array has an \ac{IEG} of 0 dB. This is expected, as it represents the gain of the reference array relative to itself. On the contrary, for the dipole arrays, the \acp{IEG} vary. Variations are larger for $\lambda/2$-spaced arrays than for $2\lambda$-spaced arrays. At large spacings, the \ac{MC} becomes negligible, and hence the \acp{EEP} become approximately equal to isolated dipole patterns, which are omni-directional with a gain of 2.15 dBi. At small spacings, however, the \ac{MC} shapes the \acp{EEP} such that the gain towards a certain \ac{UE} depends on its \ac{AOA}. The cosine elements achieve the highest \acp{IEG} for both inter-element spacings. However, they also come with the largest variations inherent to their directive element patterns. 


    The second subplots of Fig.~\ref{fig:rslt_no_dropping} and Fig.~\ref{fig:rslt_dropping} show, in dB-scale, the \acp{CDF} of the \ac{ZF} \ac{BCC} $|\omega_{kk}|^2$ \eqref{eq:SINR_ZF_SC}. Contrary to what was observed for $G^\mathrm{IE}_k$, user dropping has a significant impact on the statistics of $|\omega_{kk}|^2$ of the served \acp{UE}: it reduces the variation drastically. Moreover, it is observed that for the considered array antennas, $|\omega_{kk}|^2$ is determined to a great extent by the array layout, whereas the element type has only a small effect. Finally, it is observed that in the scenario without user dropping, the $2\lambda$-spaced \acp{NULA} significantly reduce the probability of having a low \ac{BCC}. The same arrays also provide a lower probability of dropping users. In the case of \ac{ZF} combining, a high \ac{GSCC} $|\bar{\alpha}_{kk}|^2$, and thus a low \ac{BCC} $|\omega_{kk}|^2$ \eqref{eq:omega_ZF}, implies that suppressing the intra-cell interference of the $k$\textsuperscript{th} \ac{UE} causes the $k$\textsuperscript{th} \ac{UE} itself to be suppressed as well, ultimately resulting in low \acp{SINR}. To reduce the probability of having high \acp{GSCC}, one could apply scheduling (here, user dropping). However, as can be concluded from Fig.~\ref{fig:results}, one could also exploit the array layout, thereby reducing the dropping probability. 
        

     The third subplots show the \acp{CDF} of $\mathrm{SINR}_k^\mathrm{UL}|_\mathrm{ZF}$. As expected, they show great correspondence to the individual \acp{CDF} of the \ac{IEG} and the \ac{BCC}. The resulting \ac{UE} rates $R_k^\mathrm{UL}$ are presented in the fourth subplots for $P_\mathrm{UL}=10$ dB. The dots indicate the average \ac{UE} rates $E\{R_k^\mathrm{UL}\}$ (computed with the \ac{UE} rate of a dropped \ac{UE} set to 0), such that the ergodic sum rate is found through multiplication by $K$. From the arrays considered in this work, the $2\lambda$-spaced \acp{NULA} achieve the highest ergodic (sum) rate in the scenario without user dropping. Since the UE rate is a concave function of the SINR (Fig.~\ref{fig:alpha2R}), it intuitively follows that arrays providing a low probability of low $\mathrm{SINR}_k^\mathrm{UL}$ benefit the ergodic sum rate. Although the latter can also be accomplished by employing signal processing (here, user dropping), Fig.~\ref{fig:rslt_dropping} shows that \acp{NULA} are still beneficial since they reduce the probability of a user being dropped. Since the cosine element arrays provide the largest \acp{IEG}, the $2\lambda$-spaced \ac{NULA} of cosine elements can be considered the optimal array from the ones considered here.     
     

\section{Conclusions and Future Work}
    It has been shown that \ac{SINR}-dependent \ac{QoS}-based array design in single-cell systems is a matter of shaping the probability distributions of two \acp{RV}, i.e., the \ac{IEG} and the \ac{BCC}. The concept is illustrated in detail for a \ac{FD} \ac{ZF} system, for which the latter is merely a function of the \ac{GSCC}. It is shown that ergodic sum rate enhancements reported for unconventional array layouts mainly result from a reduced probability of a high \ac{GSCC} and that such arrays can reduce the need for scheduling. In the future, we plan to use the presented concepts to design new array layouts rather than analyzing existing ones.

\begin{acks}

\end{acks}

\balance

\bibliography{IEEEabrv,iet-ell}
\bibliographystyle{iet}

\end{document}